\documentclass [12pt]{article}

\begin{document}

\begin{center}
\textbf{Fermions in conformally invariant 
geometrodynamics }
\end{center}

\begin{center}
M.V. Gorbatenko
\end{center}

\begin{center}
Russian Federal Nuclear Center - All-Russian Research Institute of\\
Experimental Physics; 37 Mir Ave., Sarov, Nizhni Novgorod Region, 607190, 
Russia; e-mail: \underline {gorbatenko@vniief.ru}
\end{center}

\subsection*{Abstract}

\textit{Dynamic equations that are the simplest conformally invariant 
generalization of Einstein equations with cosmological term are considered. 
Dimensions and Weyl weights of the additional geometrical fields (the vector 
and the antisymmetric tensor) appearing in the scheme are such, that they 
admit an unexpected interpretation. It is proved that the fields can be 
interpreted as observed, generated by bispinor degrees of freedom. The 
vacuum polarization density matrix leads to different probabilities of 
different helicity particle generation.} 

\vskip8mm

This paper leans on ref. [1] which proves the below statements and gives 
needed references.

The simplest conformally invariant generalization of Einstein equations with 
a nonzero lambda term are the following equations:
\begin{equation}
\label{eq1}
R_{\alpha \beta}  - \frac{{1}}{{2}}g_{\alpha \beta}  R = - 2A_{\alpha}  
A_{\beta}  - g_{\alpha \beta}  A^{2} - 2g_{\alpha \beta}  A^{\nu} {}_{;\nu}  
+ A_{\alpha ;\beta}  + A_{\beta ;\alpha}  + \lambda g_{\alpha \beta}  \quad 
.
\end{equation}
Equations (\ref{eq1}) retain their form in conformal transformations of field 
quantities:
\begin{equation}
\label{eq2}
\left. {\begin{array}{l}
 {g_{\alpha \beta}  \to {g}'_{\alpha \beta}  = g_{\alpha \beta}  \cdot 
exp\left[ {2\phi \left( {x} \right)} \right],} \\ 
 {A_{\alpha}  \to {A}'_{\alpha}  = A_{\alpha}  - \frac{{\partial \phi \left( 
{x} \right)}}{{\partial x^{\alpha} }},} \\ 
 {\lambda \to {\lambda} ' = \lambda \cdot exp\left[ { - 2\phi \left( {x} 
\right)} \right],} \\ 
 \end{array}}  \right\}
\end{equation}
\noindent
where $\phi \left( {x} \right)$ is an arbitrary function of coordinates. 

Equations (\ref{eq1}) can be considered as a extension of Einstein equations to Weyl 
space. In the Weyl space each geometrical quantity is assigned the 
characteristic, such as Weyl weight $n$, which is an integer. If a quantity 
has weight $n$, then in transformations (\ref{eq2}) it is multiplied by
 $exp\left[ {n 
\cdot \phi \left( {x} \right)} \right]$. 

From the above-said and from (\ref{eq2}) it follows that the weight of the metric 
tensor is $n = - 2$ and that of the lambda term is $n = + 2$. Vector 
$J^{\alpha} $ and antisymmetric tensor $H^{\alpha \beta} $, which will play 
a specific role in what follows, are:
\begin{equation}
\label{eq3}
J^{\alpha}  \equiv \frac{{1}}{{\lambda ^{3/2}}}g^{\alpha \beta} \left( 
{\lambda _{;\beta}  - 2A_{\beta}  \lambda}  \right),
\end{equation}
\begin{equation}
\label{eq4}
H^{\alpha \beta}  \equiv \frac{{1}}{{\lambda} }g^{\alpha \mu} g^{\beta \nu 
}\left( {A_{\nu ;\mu}  - A_{\mu ;\nu} }  \right).
\end{equation}
From equations (\ref{eq1}) it follows that
\begin{equation}
\label{eq5}
\left( {\lambda H_{\alpha}  \;^{\nu} } \right)_{;\nu}  = \lambda 
^{3/2}J_{\alpha}  ,
\end{equation}
\begin{equation}
\label{eq6}
\left( {\lambda ^{3/2}J^{\alpha} } \right)_{;\alpha}  = 0.
\end{equation}

For the following consideration, definition of bispinor matrix $Z$ will be 
needed. Like in Riemannian space, in Weyl space this matrix is introduced 
through world and local Dirac matrices $\gamma _{\alpha}  $, $\gamma _{k} $ 
with relation
\[
\gamma _{\alpha}  \left( {x} \right) = H_{\alpha} ^{k} \left( {x} \right) 
\cdot Z^{ - 1}\left( {x} \right)\gamma _{k} Z\left( {x} \right).
\]
Here $H_{\alpha} ^{k} \left( {x} \right)$ is a system of reference vectors 
determined by relation $g_{\alpha \beta}  = H_{\alpha} ^{m} H_{\beta} ^{n} 
g_{mn} $. 

Particular role of (\ref{eq3}), (\ref{eq4}) is that it is only they that can be used to 
construct Hermitean matrix
\begin{equation}
\label{eq7}
M = J^{k} \cdot \left( {\gamma _{k} D^{ - 1}} \right) + H^{mn} \cdot \left( 
{S_{mn} D^{ - 1}} \right),
\end{equation}
\noindent
all parts of which are dimensionless and have zero Weyl weight.

The tensor system which can be mapped onto bispinors in the 4-dimensional 
Weyl space is composed of a scalar, a vector, an antisymmetric tensor, a 
pseudo-vector, and a pseudo-scalar. However, in the case under discussion, 
among the system only vector (\ref{eq3}) and tensor (\ref{eq4}) are suitable for the 
purposes of the mapping onto bispinors. 

Matrix (\ref{eq7}) is Hermitean. If it is, in addition, positive, then find $H$ - 
arithmetic root of it.
\begin{equation}
\label{eq8}
M\left( {x} \right) = H\left( {x} \right)H^{ +} \left( {x} \right).
\end{equation}

The square root extraction operation with the requirement of the hermiticity 
and the root positivity is unique, if matrix $M$ is positive. Thus, equation 
(\ref{eq8}) is an algebraic equation for finding $H$ by Cauchy data for equation 
(\ref{eq1}). Recall that for equations (\ref{eq1}) the Cauchy problem is well-posed without 
any connections to the Cauchy data on the initial hypersurface.

Write the polar decomposition for the bispinor matrix as
\begin{equation}
\label{eq9}
Z = HU^{ - 1}.
\end{equation}
With taking into account (\ref{eq11}) we obtain
\begin{equation}
\label{eq10}
M\left( {x} \right) = H\left( {x} \right)H^{ +} \left( {x} \right) = Z\left( 
{x} \right)Z^{ +} \left( {x} \right).
\end{equation}
Vector (\ref{eq3}) and anti-symmetric tensor (\ref{eq4}) are expressed in terms of the 
bispinor matrix as follows:
\begin{equation}
\label{eq11}
J^{\alpha}  = {\textstyle{{1} \over {4}}}Sp\left[ {Z^{ +} C\gamma _{5} 
\gamma ^{\alpha} Z} \right],
\quad
H^{\alpha \beta}  = - {\textstyle{{1} \over {8}}}Sp\left[ {Z^{ +} CS^{\alpha 
\beta} Z} \right].
\end{equation}

Thus, in the scheme under consideration the Hermitean multiplier $H\left( 
{x} \right)$ in (\ref{eq8}) is determined by equations (\ref{eq1}). As for the unitary 
multiplier $U^{ - 1}\left( {x} \right)$, the simplest conformally invariant 
equation for it is
\begin{equation}
\label{eq12}
\left( {HD\gamma ^{\alpha} H} \right)\left( {\left( {\nabla _{\alpha}  U^{ - 
1}} \right)U} \right) = - {\textstyle{{1} \over {2}}}\left( {\nabla _{\alpha 
} \left( {HD\gamma ^{\alpha} H} \right)} \right) - {\textstyle{{3} \over 
{4}}}\left( {HD\gamma ^{\alpha} H} \right) \cdot \left( {ln\lambda}  
\right)_{;\alpha}  ;
\end{equation}

Introduce the following notation for the anti-Hermitean matrix vector $iZ^{ 
+} D\gamma ^{\alpha} Z$:
\begin{equation}
\label{eq13}
\Upsilon ^{\alpha}  \equiv iZ^{ +} D\gamma ^{\alpha} Z.
\end{equation}
With using notation (\ref{eq13}), equation (\ref{eq12}) assumes the following compact form:
\begin{equation}
\label{eq14}
\left( {\nabla _{\alpha}  \left( {\lambda ^{3/2} \cdot \Upsilon ^{\alpha} } 
\right)} \right) = 0.
\end{equation}
For matrix connectivity $\Gamma _{\alpha}  $, determine the matrix curvature 
tensor (Young-Mills tensor in conventional terms) in the usual manner: 
\[
P_{\alpha \beta}  \equiv \Gamma _{\beta ;\alpha}  - \Gamma _{\alpha ;\beta}  
+ \Gamma _{\alpha}  \Gamma _{\beta}  - \Gamma _{\beta}  \Gamma _{\alpha}  .
\]
The only possible form of the conformally invariant Young-Mills equation is
\begin{equation}
\label{eq15}
\left( {\nabla _{\nu}  P_{\alpha}  \;^{\nu} } \right) = Const \cdot \left( 
{\lambda ^{3/2} \cdot \Upsilon _{\alpha} }  \right).
\end{equation}
Relation (\ref{eq14}) ensures consistency of the left-hand and right-hand sides in 
equation (\ref{eq15}).

From the method the matrix $M$ is introduced with it follows that it has the 
meaning of the polarization density matrix. In our case there are two types 
of bispinor degrees of freedom. The degrees relating to the Hermitean 
multiplier determine the matrix $M$ itself, therefore, the matrix $M$ can 
not be a polarization matrix with respect to them. As for the degrees 
relating to the unitary multiplier $U^{ - 1}$, the matrix $M$ appears in the 
dynamic equations for $U^{ - 1}$. As $U^{ - 1}$-related degrees of freedom 
are external, ``nested'' bispinor degrees of freedom, then, when treating 
their dynamics, the matrix $M$ should be considered as the one determining 
polarization properties of background vacuum. The quantity $M$ (and thereby 
the quantities $A_{\alpha}  $ and $\lambda $) determine relative 
probabilities of generation of virtual particles with half-integer spin in 
different states. Thus, the physical meaning of the quantities $A_{\alpha}  
$ and $\lambda $ is that they give properties of the geometrodynamical 
vacuum, which is background vacuum for other degrees of freedom (that is for 
the bispinor degrees determined by the unitary multiplier $U$ and for the 
Young-Mills fields). 

Presume that the polarization matrix $M$ is an operator in the space of 
half-integer spin particle states. That is make an assumption, which is 
typically made in quantization of the classic field theory apparatus. Then 
the case of $M = E$ will describe the situation, where virtual particles 
with half-integer spin can be generated with the same probability in any of 
possible states.

The question arises: How should matrix $M$, whose eigenvalues are not equal 
to one, be treated? A possible treatment is that in the considered space the 
probabilities of generation of virtual bispinor particles in one state are 
somewhat higher than those in another state. In the theory of matrix spaces 
bispinors are separated from the bispinor matrix through multiplication of 
$Z$ by projector on the left. In the general case the projector can be 
produced with using matrix $i\gamma _{5} $ and/or matrix of mapping from 
group $U_{4} $. Among these two matrices the state partition into $2 + 2$ 
can be achieved only with the $i\gamma _{5} $ separating the states with the 
left and right helicity. So in the scheme under discussion the virtual 
bispinor particles of different helicity have different probabilities of 
generation.

It is not impossible that the above mechanism leads to appearance of baryon 
asymmetry in the Universe. A clue to the puzzle A.D. Sakharov was concerned 
about - What is the reason for the baryon asymmetry in the Universe? - may 
be along this path. 

The above considerations are no more than hypotheses and require further 
studies. Quite reliable seems the relationship found in this paper between 
the tensor field dynamics and the bispinor degree of freedom dynamics. Of 
quite general nature is also the division of the bispinor degrees of freedom 
into two parts: (\ref{eq1}) the part contained in $H$ (Hermitean multiplier in the 
bispinor matrix factorization) and rigidly connected to the background 
space-time vacuum; (\ref{eq2}) the part contained in $U^{ - 1}$ (unitary multiplier 
in the bispinor matrix factorization) and having (along with the matrix 
connectivity) the meaning of a field ``nested'' in a given geometrodynamical 
background vacuum.

As far as we know, the analysis method used is novel and may be useful for 
interpretation not only of the scheme based on equation (\ref{eq1}), but also of 
other theories aimed at a unified description of fundamental interactions.

\section*{References}

[1] M.V. Gorbatenko. \textit{Voprosy Atomnoi Nauki i Tekhniki}. Seriya: 
Teor. i Prikl. Fizika. \textbf{3}, 28-40 (2001) [In Russian].

\end{document}